\documentclass[twocolumn,prb,aps,footinbib,amsmath,amssymb,superscriptaddress]{revtex4-1}
\usepackage{graphicx}
\usepackage{dcolumn}
\usepackage{xcolor}
\usepackage{bm}
\usepackage{natbib,hyperref}
\usepackage{amsmath,amssymb,amsfonts,bbold}
\usepackage[normalem]{ulem}
\usepackage{dirtytalk}

\begin{document}

\title{In-gap states induced by magnetic impurities on wide-band s-wave superconductors: self-consistent calculations}
\author{Divya Jyoti}
\affiliation{Centro de F{\'{i}}sica de Materiales
        CFM/MPC (CSIC-UPV/EHU),  20018 Donostia-San Sebasti\'an, Spain}
\affiliation{Donostia International Physics Center (DIPC),  20018 Donostia-San Sebasti\'an, Spain}    
\author{Deung-Jang Choi}
\affiliation{Centro de F{\'{i}}sica de Materiales
        CFM/MPC (CSIC-UPV/EHU),  20018 Donostia-San Sebasti\'an, Spain}
\affiliation{Donostia International Physics Center (DIPC),  20018 Donostia-San Sebasti\'an, Spain}
\author{Nicol{\'a}s Lorente}
\email{nicolas.lorente@ehu.eus}
\affiliation{Centro de F{\'{i}}sica de Materiales
        CFM/MPC (CSIC-UPV/EHU),  20018 Donostia-San Sebasti\'an, Spain}
\affiliation{Donostia International Physics Center (DIPC),  20018 Donostia-San Sebasti\'an, Spain}

\begin{abstract}
The role of self-consistency in Bogoliubov-de Gennes equations is frequently underestimated in the investigation of in-gap states created by magnetic impurities in s-wave superconductors. Our research focuses on the impact of self-consistency on the in-gap states produced by magnetic stuctures on superconductors, specifically evaluating the density of states, the in-gap bands, and their topological attributes. 
Here, we show results ranging from single impurity to finite chains, and infinite ferromagnetic spin chains in wide-band s-wave superconductors.
These results show that the order parameter contains important information regarding quantum phase transitions and their topological nature, underscoring the importance of self-consistency in such studies. 

\end{abstract}
\date{\today}

\maketitle
\section{Introduction}
The Bogoliubov-de Gennes (BdG) equations have played a pivotal role in simplifying and advancing the theoretical examination of magnetic impurities within superconductors. The method and examples have been clearly exposed in Refs.~\onlinecite{Zhu_2006,Zhu_2016}. The BdG framework offers a means to address the complex many-body problem associated with superconductivity by employing a mean-field approximation of the pairing interaction. The final equations amount to a diagonalization of a one-body effective Hamiltonian~\cite{Zhu_2006}. It is essential to recognize that the mean-field nature of the BdG equations necessitates the self-consistent evaluation of the effective pairing interaction.

In a superconductor, a single magnetic impurity leads to weakening of the Cooper pair binding energy. This can give rise to the appearance of one-particle states in the superconducting gap~\cite{YuLuh_1965,Shiba_1968,Rusinov_1969, Zhu_2006}. 
Approximating the impurity-substrate interaction by the Kondo Hamiltonian \cite{Schrieffer_Wolff}, the minimum energy required for the emergence of quasiparticle excitations decreases when the Kondo-exchange coupling interaction, $J$ increases. 
Near a critical value of the exchange interaction ($J_c$), the superconducting condensate becomes thermodynamically unstable against the formation of  quasiparticles~\cite{Sakurai_1970,Salkola_1997,Sakurai_2013}. Henceforth, a superconductor undergoes a quantum phase transition (QPT) and achieves a spin-polarized state. Theoretical studies \cite{Salkola_1997, Flatte_1997a, Flatte_1997b, Flatte_2000, Hoffman_2015, Bjornson_2017, Theiler_2019, Cuevas_2020} and experimental realizations of QPT have been reported by several groups \cite{Hatter_2015, Farinacci_2018, Huang_2020, Liebhaber_2021, Karan_2022, Liu_2022, Zhou_2022, Uldemolins_2023} where a range of Kondo-exchange couplings are achieved based on how adatoms attach to specific sites on the substrate \cite{Hatter_2015,Franke_2011, Hua_2010}. Alternatively, by establishing an electrical control in Josephson junction systems \cite{Maurand_2012, Delagrange_2015, Delagrange_2016, Deacon_2010} or by changing tip-sample distances in scanning tunneling microscopy (STM) experiments \cite{Farinacci_2018}, these couplings can be tuned. The strength and nature of such couplings can have a direct impact on the superconducting state and therefore incorporating self-consistency may be necessary to obtain an accurate description of such systems.

Many calculations have demonstrated that the influence of self-consistency is predominantly quantitative, suggesting that the underlying physics can be accurately captured using simpler and more computationally efficient non self-consistent approaches~\cite{VonOppen_2021, Salkola_1997,Flatte_1997a,Paaske_2016,Karan_2022}. However, there exist specific scenarios where the adoption of self-consistent solutions is essential~\cite{Zonda_2015, Rozhkov_1999}. Levy Yeyati et al. ~\cite{Yeyati_1995} illustrated that concerns regarding the violation of particle-number conservation in the BdG formalism are resolved when self-consistency is incorporated in electric current evaluations. Particularly critical are instances involving QPTs, where the interaction between impurity and superconductor is so substantial that it alters the ground state of the entire system, indicating a zero-temperature QPT and underscoring the significance of self-consistency in these analyses. Notably, the closure of the superconducting gap, a precursor to a QPT, is an example of behavior that qualitatively differs when including self-consistency \cite{Salkola_1997,Franke_2011,Hatter_2015,Heinrich2018}. This gap closure also forms a necessary criterion for the emergence of topological quantum phase transitions (TQPTs) observed in spin chains on s-wave superconductors \cite{Beenakker,Pientka_2013,Choi_2019}. Consequently, examining topological invariants and phases through self-consistent methods becomes a critical aspect when employing BdG equations \cite{Bjornson_2015,Christensen_2016,Bjornson_2017,Theiler_2019}.


In this study, we expand on our prior theoretical work on spin impurities in s-wave superconductors \cite{Mier2_2021, moire}, focusing on the self-consistent evaluation of the order parameters. Our previous research delved into topological transitions in wide-band s-wave superconductors \cite{Mier2_2021, moire}, relating them to experimentally observed non-zero edge states \cite{Schneider2}. In our current work, we utilize self-consistent solutions not only to validate and reinforce the findings of our earlier studies but also to deepen our comprehension of QPTs. This approach is inspired by literature \cite{Bjornson_2015, Bjornson_2017, Theiler_2019} that underscores the significance of self-consistency in determining the topological properties of spin chains on superconductors. However, contrasting views, especially from Ref. \onlinecite{Christensen_2016}, suggest a more indirect role of self-consistency, impacting the phenomena through renormalization of critical exchange interactions and consequent gap closure, leading to QPTs.
 
The structure of our paper is as follows: Initially, we present our numerical method, based on the techniques developed by Flatté and Byers~\cite{Flatte_1997a, Flatte_1997b} in a discretized form~\cite{Mier2_2021}. We first test the correctness of the approach by retrieving the BCS temperature dependence of a bulk wide-band superconductor. Next, we add a magnetic impurity and recover the properties of the induced in-gap states. Subsequently, we conduct a systematic investigation into the impact of progressively increasing the number of impurities on a superconductor, leading to an in-depth analysis of ferromagnetic infinite chains and their associated topological invariants. We show that the use of the order parameter as a diagnostic tool proves to be exceptionally insightful for detecting the emergence of QPTs, particularly in the realm of TQPTs. A principal finding of our research is that self-consistency does not alter the topological properties previously identified in non-self-consistent studies and has a minimal effect on the density of states in realistic wide-band s-wave superconductors. 

\section{Theoretical methodology}

\subsection{Bogoliubov-de Gennes equations using Green's functions}
We solve the BdG using the Nambu formalism and a discretization of the spatial continuum\cite{Pientka_2013,Mier_2021,Cuevas_1996}. We start from the Hamiltonian for a pristine BCS superconductor given by,
\begin{equation}
    \Hat{H}_{BCS} = \xi_k\tau_0\sigma_3 + \Delta \tau_2\sigma_2,
    \label{BCS}
\end{equation}
where $\sigma_i$ ($\tau_i$) are the Pauli matrices acting on the
spin (particle) subspace, $\xi_k$ is the energy from the Fermi level 
($\xi_k = \epsilon_k - E_F$) and $\Delta$ is the superconducting gap or order parameter. The previous Hamiltonian is written in the
4-dimensional Nambu basis: $\Psi = (\psi_{\uparrow}, \psi_{\downarrow},
\psi_{\uparrow}^\dagger, \psi_{\downarrow}^\dagger)^T$.

To model the experimental system, we add the Hamiltonian describing the magnetic impurities~\cite{Flatte_2000, Flatte_1997a,Flatte_1997b}, but assuming strictly localized interactions~\cite{Pientka_2013}.
\begin{equation}
 \Hat{H}=   \Hat{H}_{BCS} + \Hat{H}_{impurity} =\Hat{H}_{BCS} + \sum_j^N (U_j \tau_3+J_j\Vec{S_j}\cdot\Vec{\alpha})\; 
 \label{Ham}
\end{equation}
with $\Vec{\alpha} = \frac{1 + \tau_3}{2}\Vec{\sigma} + \frac{1 -
\tau_3}{2}\sigma_2\Vec{\sigma}\sigma_2$, where $\Vec{\sigma}$ is the spin
operator~\cite{Shiba_1968}.

The interaction contains an exchange coupling, with strength $J_j$, and a non-magnetic potential scattering term, $U_j$,  per impurity $j$. The impurity spin is assumed to be a classical vector, $\Vec{S_j}$, within the classical-spin approximation, see \ref{classical}.

The Hamiltonian is completed by a Rashba term:

\begin{eqnarray}
\label{Rashba}
    \Hat{H}_{Rashba}&=&i \frac{\alpha_R}{2a} \sum_{i,j,\alpha, \beta}[\hat{c}^\dagger_{i+1,j, \alpha}
(\sigma_2)_{\alpha, \beta} \hat{c}_{i,j, \beta} \nonumber \\
&-&\hat{c}^\dagger_{i,j+1, \alpha}
(\sigma_1)_{\alpha, \beta} \hat{c}_{i,j, \beta} +h.c.]
\end{eqnarray}
where ${\alpha, \beta}$ are {the spin indexes}. Here, ${i,j}$ represent the impurity index in a two-dimensional lattice. This interaction
couples spins on next-nearest-neighbour sites. The lattice parameter of the substrate
is $a$, and the factor of $2 a$ comes from a finite-difference scheme to
obtain the above discretized version of the Rashba interaction. 

We solve the BdG equations using Green's functions, 
 ${G}^{\nu,\mu}_{i,j}$ evaluated for the sites $i$ and  $j$
for the Nambu components $\nu$ and $\mu$ by solving Dyson's equation:
\begin{equation}
\label{Dyson2}
    \Hat{G} = [\Hat{G}_{BCS}^{-1} - \Hat{H}_{I}]^{-1}
\end{equation}
where $\Hat{G}_{BCS}$ is the retarded Green's operator for the BCS Hamiltonian from Eq.~(\ref{BCS}) and $\Hat{H}_I = \Hat{H}_{impurity} + \Hat{H}_{Rashba}$. In the present study, we will be interested in the effect of the impurities on the local gap $\Delta_i$, for this, we use the local Green's function.
The Green's function that is diagonal on spatial indices, $i$ and $j=i$, is given by the  usual local BCS Green's function~\cite{Vernier_2011} (following usual convention $\hbar=1$ unless otherwise specified),
\begin{eqnarray}
{G_{i,i}}_{BCS} ( \omega)
&=& - \frac{\pi N_0 Sgn [Re(\omega)Im(\omega)]}{\sqrt{\Delta^2-\omega^2}} \nonumber \\
&\times&
\begin{pmatrix}
\omega & 0 & 0 & -\Delta \\
0 & \omega &\Delta& 0\\
0 &\Delta&\omega& 0 \\
-\Delta & 0 & 0 & \omega
\end{pmatrix}.
\label{GBCS0}
\end{eqnarray}
Here, $N_0$ corresponds to the normal-metal density of electronic states evaluated at the Fermi energy.
The non-local Green's function can be approximated by a well-known analytical expression, see for example Refs.~\onlinecite{Flatte_1997a,Meng_2015,Paaske_2016,Mier2_2021}, yielding
\begin{eqnarray}  
    {G}_{BCS}(\Vec{r},\Vec{r'},\omega) &=& -\frac{\pi N_0}{k_Fr}e^{\frac{-\sqrt{\Delta^2 - \omega^2}r}{\pi\xi\Delta}}\times
(\cos(k_F r) \tau_0 \sigma_3\;
\nonumber \\
&+&
 \frac{\sin(k_F r)}{\sqrt{\Delta^2-\omega^2}}
(\omega \tau_0 \sigma_0 + \Delta \tau_2 \sigma_2)),
\label{GBCS}
\end{eqnarray}
where $r=|\Vec{r}-\Vec{r'}|$, $k_F$ is the Fermi wave-vector and $\xi=\frac{\hbar v_F}{\pi \Delta}$ is coherence length.

Within this formalism the condition for self-consistency comes from the mean-field value of the gap function~\cite{Zhu_2016}:
\begin{equation}
\Delta (\vec{r}, \vec{r'})=
V (\vec{r}- \vec{r'}) \langle\psi_\downarrow (\vec{r}) \psi_\uparrow (\vec{r'})\rangle
\label{Gap_realspace}
\end{equation}
where the unknown pairing potential $V(\vec{r}- \vec{r'})$ is approximated by a local potential. Additionally, a cutoff in energies is applied in the evaluation of the ground-state average, just taking quasi-particle states that are within the Debye frequency, $\omega_D$. The average is easy to evaluate using Green's functions, in particular we calculate:
\begin{equation}
\Delta_i = \tilde{V}
\int^{\hbar \omega_D}_{-\hbar \omega_D}
f(\omega) Im \{G^{1,4}_{i,i}(\omega)-G^{2,3}_{i.i}(\omega)\} d\omega.
\label{GapSelf}
\end{equation}
Here, $f(\omega)$ is the Fermi function at temperature $T$, evaluated for energy $\hbar \omega$, $G^{1,4}$ and $G^{2,3}$ are the anomalous Green's function components for Nambu indices 1, 4 and 2, 3, respectively. The effective interaction with all possible multiplicative constants,
$\tilde{V}$, is assumed to be homogeneous and independent of the impurity interactions. It can be easily obtained from the value $\Delta_0$ for the pristine superconductor and the BCS Green's function, Eq. (\ref{GBCS0}). See appendix~\ref{AppendixV}.

The new $\Delta$ is computed by using the Green's functions solving Dyson's equations for the full system. The change in the order parameter, $\delta \Delta (\vec{r})$ is then used as a self-energy to compute the new inhomogeneous Green's function following Refs. \onlinecite{Flatte_1997a,Flatte_1997b}. This procedure is repeated
until $\Delta(\vec{r})$ changes less than a certain tolerance.

Now for the evaluation of infinite chains, we rewrite the Green's functions in terms of their k-space values. For this, we take into account that
\[ G_{i,j} (\omega)= \frac{1}{N_k}\sum_k G(k, \omega) e^{i k(R_i-R_j)},\]
where $N_k$ refers to the number of k-points of the calculation and $R_i, R_j$ describes discrete position coordinates.

Then from Eq. (\ref{GapSelf}), we obtain
\begin{eqnarray}
\Delta_i = \frac{\tilde{V} }{N_k}\int^{\hbar \omega_D}_{-\hbar \omega_D}f(\omega)  \nonumber 
\sum_k Im \{G^{1,4}(k,\omega) - \\ G^{2,3} (k,\omega)\} d\omega , 
\label{Gap_kspace}
\end{eqnarray}
where $\tilde{V} (\vec{r}-\vec{r}')$ is assumed homogeneous in an isotropic s-wave type superconductor, and constant. Since we are representing an infinite and homogeneous system, the above order parameter $\Delta_i$ does not depend on the site $i$ and is constant. We can actually obtain the k-resolved expression by removing the sum over k points in Eq. (\ref{Gap_kspace}), this yields
\begin{eqnarray}
\Delta_k = \tilde{V}\int^{\hbar \omega_D}_{-\hbar \omega_D}f(\omega)  \nonumber 
 Im \{G^{1,4}(k,\omega) - \\ G^{2,3} (k,\omega)\} d\omega , 
\label{Gap_kkspace}
\end{eqnarray}
such that the average order parameter is $\Delta_i =\frac{1}{N_k}\sum_k\Delta_k$.
The advantage of the above expression, Eq. (\ref{Gap_kspace}) is that the k-space Green's functions can be easily evaluated~\cite{Peng_2015,Paaske_2016}.

The self-consistent calculation for the infinite chain is performed by computing $\tilde{V}$ using the homogeneous superconductor. This means setting all interactions to zero except the Rashba coupling. Then the value of $\Delta_i$ fixes the coupling  $\tilde{V}$. Next, we proceed to define a zero-iteration $\Delta_k$. This is done by evaluating Eq. (\ref{Gap_kkspace}) for the free Green's functions, i.e. without interactions and only the Rashba coupling. Next, the interactions are included using Dyson's equations and the above self-consistent loop is performed by defining a new self-energy that contains the change of $\Delta_k$ per k-point and per iteration.

\subsection{Numerical implementation}\label{Numerical}

We use a lattice approach discretizing the spatial dependence. The size of the discrete lattice step, $\Delta r$, is quite robust against different values. 
The approximations leading to  the analytical BCS Green's function, imply that $\Delta r \times k_F >1$, where $k_F$ is the Fermi wave vector. 
Additionally, special care must be paid in converging
the Rashba interaction, due to the evaluation of the Rashba gradients on the grid. Yet, the robustness of the full approach comes from using a $G_{BCS}$ that is defined everywhere in space. Here, we use 3-D Green's functions because using 2-D Green's functions just bring small numerical changes \cite{Brydon_2015} despite the different in-gap state decay~\cite{Menard_2015}, that does not affect the convergence of the order parameter, the main object of our present study.

A consequence of these approximations for the real-space  BCS Green's function is that $G_{BCS}(k, \omega)$ is defined
only for $k < k_F$. As a consequence, the Brillouin-zone $\pi/a$ where $a$ is the distance between impurities, should be smaller than $k_F$ to have the Green's function defined for all k-points. A simple compromise is to take $\Delta r\approx a$ for a large enough $k_F$, please see the corresponding
discussion in Ref.~\onlinecite{Mier2_2021}. As in Ref.~\onlinecite{Pientka_2013,Schneider1}, our calculations are best suited for spin chains in the dilute-impurity limit.

\subsection{Classical-spin approximation}\label{classical}

The classical-spin approximation was used by Yu, Shiba and Rusinov \cite{YuLuh_1965,Shiba_1968,Rusinov_1969} to prove and characterize the appearance of in-gap states in the presence of a magnetic impurity. However, it was soon recognized that the impurity spin is a quantum operator and spin fluctuations occur that can eventually produce the Kondo effect \cite{Matsuura}. In addition to spin fluctuations, spin-orbit coupling can lift the degeneracy of the impurity's spin states, presenting different states that can be accessible at different energies leading to a different picture of the low-energy physics \cite{Zitko_2011,VonOppen_2022}.

The classical spin approximation gives sensible results in the large-spin limit given by $S_i\rightarrow\infty$ with $J\times S_i$ finite. Reference \onlinecite{Zitko_2011} shows that for transition-metal spins, this limit is not satisfied. And despite the quenching of spin fluctuations by the presence of the superconducting gap, at $J\times S_i$ large enough to yield bound in-gap states, spin fluctuations become available again.

In the present work, we focus on the properties of in-gap states without studying the nature of the many-body ground state~\cite{Moca_2008} or spin transitions in the impurity~\cite{VonOppen_2021}. In these conditions, the classical-spin approximation yields a sufficient description that is easily accessible with mean-field theories such as the BdG approach used here.

\section{Temperature effects}

\begin{figure} [t!]
    \begin{center}
    \includegraphics[width = 0.35\textwidth]{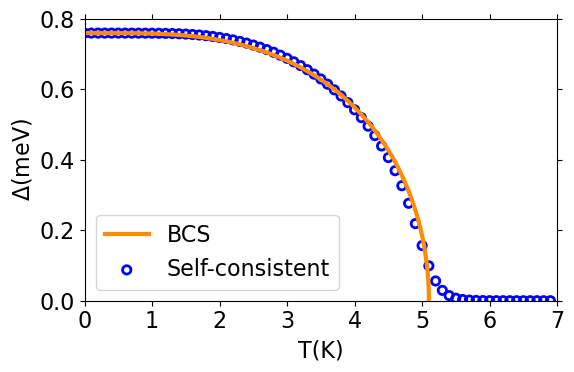}
    \end{center}   
\caption{Self-consistently evaluated superconducting gap, $\Delta$ computed as a function of temperature for a BCS model of an s-wave superconductor, Bi$_2$Pd. A BCS curve following Eq.~(\ref{BCS_curve}) is also shown for comparison. To obtain a critical temperature  $T_c = 5.6$~K (in good agreement with the experimental $T_c=5.4$~K \cite{Imai_2012}), we only used $\Delta_0= 0.76$ meV and the Bi$_2$Pd Fermi wave vector, $k_F=0.183~a_0^{-1}$, as inputs.}
\label{BCSgap}
\end{figure}

The self-consistent BdG approach readily captures the well-established BCS behavior of the superconducting gap as a function of temperature. In Fig.~\ref{BCSgap}, we present the outcomes for the Bi$_2$Pd s-wave superconductor, utilizing our free-electron BdG equations to simulate the BCS superconductor. To enhance comparability, we include a BCS curve, derived from the subsequent expression,
\begin{equation}
    \Delta(T) = \Delta_{0} \tanh \Biggl(\ 1.74 \sqrt{\frac{T_c}{T}-1}\Biggl).
    \label{BCS_curve}
\end{equation}
From Fig.~\ref{BCSgap}, we obtain a critical temperature of 5.6~K, which is in good agreement with the experimental one $T_c=5.4$ K as reported in Ref.~\onlinecite{Imai_2012}. To obtain this value, the $T=0$ value of the gap, $\Delta_0$, has been fixed to the experimental one, 0.76~meV. The only other parameter is the free-electron density fixed by the Fermi wave vector $k_F$, that has been taken as 0.183~$a_0^{-1}$ following the experimental data of Ref.~\onlinecite{Herrera_2015}. The results are largely independent of the Dynes parameter used as imaginary part of $\omega$ in the Green's function expressions \cite{Dynes_1978}. For the present calculation a Dynes parameter of $\Gamma=0.05$ meV was used.

\section{Single magnetic impurity on a superconductor}

Let us first study a single magnetic impurity using the above formalism. Here, we will briefly characterize the in-gap state as with respect to its extension in the superconductor and to the electron-hole symmetry built-in the equations.

\subsection{Extension of the in-gap states}\label{extension}

In simple arguments, there are two different length scales associated to a single Shiba state~\cite{Salkola_1997,Flatte_1997b}. The longer scale is due to the actual extension of correlations in the superconductor. It is related to the coherence length, $\xi$, given by the decay length $\xi'=\xi/ \sqrt{1-\frac{\omega^2}{\Delta^2}}$, where $\omega$ is the pole of the Shiba state or the quasienergy of the Shiba excitation.

The second length scale is the immediate decay of the Shiba wavefunction as soon as we move away from the impurity center. Our model is particularly unrealistic on the impurity site due to the presence of a $\delta$-function in the interaction that describes a very localized magnetic interaction (Eq. \ref{GBCS0}). Within this approximation, we see that the Shiba wave function has a decay that is actually given by $(\lambda_F/r)^2$, where the Fermi wavelength is given by $\lambda_F=2\pi/k_F$ and $r$ is the spatial distance to the impurity's center.

We can easily retrieve this behavior from our theoretical model. The wavefunction can be obtained from the Green's function using the Lippmann-Schwinger equation:
\begin{equation}
    \psi(\vec{r})=\psi_0 (\vec{r}) + \int d\vec{r'}
    G_{BCS}(\vec{r},\vec{r'},\omega) H_{impurity} (\vec{r'})
    \psi(\vec{r'}).
    \label{Lipp}
\end{equation}

For Shiba states, $\psi_0 (\vec{r})=0$ because there are no single-quasiparticle states in the gap of the pristine superconductor. Using the locality of the  interaction with the impurity, $H_{impurity} (\vec{r}) \propto \delta (\vec{r})$, we easily simplify Eq. (\ref{Lipp}) to 
\begin{equation}
    \psi(\vec{r}) \propto
    G_{BCS}(\vec{r},0,\omega)
    \psi(0)
    \label{Lipp2},
\end{equation}
where $G_{BCS}(\vec{r},0,\omega)$ can be calculated from Eq. (\ref{GBCS}).

Then, away from the impurity's center the wavefunction decays as the bare BCS Green's function. Using the typical BCS expression for the Green's function~\cite{Flatte_1997a,Meng_2015,Paaske_2016,Mier2_2021}, we see that this decay is of the form $(\lambda_F/r) \, \exp(-r/\xi')$. The density scales as the square of the wave function and hence, as $(\lambda_F/r)^2 \, \exp(-2r/\xi')$ in agreement with the discussions found in Refs.~\onlinecite{Salkola_1997,Flatte_1997b,Menard_2015}. For a truly 2-D system, the decay of the Shiba states would rather scale as $1/r$, following
the arguments of Ref.~\onlinecite{Menard_2015}.

Then, two length scales are expected for the Shiba states found inside a superconducting gap. The shorter length scale of the Shiba state is then related to the inverse of the Fermi wave vector. In BCS superconductors, we can expect that larger electron densities (larger $k_F$) leads to short-ranged Shiba states. This scale is the one responsible for the hybridization of Shiba states and the formations of bands needed for the appearance of a topological phase. Then, impurities on less dense superconductors, can be placed at larger distances and still create a rich structure of in-gap bands. Indeed, Ref.~\onlinecite{Mier_2021} shows that Majoranas can be found at the edges of spin chains under these conditions, that perfectly match Cr chains on the s-wave superconductor Bi$_2$Pd.

\subsection{Particle-hole symmetry for in-gap states}

Precise experiments carried on with the STM have shown that in-gap states are not electron-hole symmetric \cite{Ruby_2016,Choi_2017}. Indeed, despite having an electron and a hole component, the in-gap states show very different spatial distributions for the electron and hole components at the exact same energies (with opposite sign).
However,
BdG equations are electron-hole symmetric. This means~\cite{Zhu_2016} that for every solution at energy $E$, with wavefunction ${\phi_i} =     (u_{i\uparrow}, u_{i\downarrow}, v_{i\uparrow},  v_{i\downarrow})^T  $, there is a solution at energy $-E$ with wavefunction ${\phi_i}' =     (-v^*_{i\uparrow}, u^*_{i\downarrow}, v^*_{i\downarrow},  -u^*_{i\uparrow})^T$,where
the electron, $u_\uparrow$, and hole, $v_\downarrow$, components of the wave functions are in principle different as experimentally found \cite{Ruby_2016,Choi_2017}. 

 Before the experimental realizations, Ref.~\onlinecite{Flatte_1997b} showed ample evidence of cases where the electron and hole components of Shiba states are not equal in magnitude. The authors of  Ref.~\onlinecite{Flatte_1997a, Flatte_1997b, Yazdani_1997} argue that even for $U_i=0$ in Eq.~(\ref{Ham}) the electron and hole components can have different magnitudes. This statement is contrary to what is stated in Ref.~\onlinecite{Salkola_1997} and contrary to what is widely accepted that \textit{electron-hole symmetry} is only attained for $U_i=0$.

As argued in Ref.~\onlinecite{Flatte_1997b}, the reason behind the $U_i=0$ electron-hole symmetry is due to the local $\delta$-like approximation for the impurity interactions. It is easy to see that this is indeed, the case from Eq. (\ref{GBCS0}) that a local potential will not lift the electron-hole symmetry at the impurity site. However, as long as the Green's function is not evaluated for the same site ($\vec{r}\neq\vec{r}'$), the electron-hole symmetry is lifted. Indeed, the particle component in Eq. (\ref{Lipp2}) depends on
\[
\cos (k_F r) + \frac{\omega}{\sqrt{\Delta_0^2-\omega^2}} \sin (k_F r),
\]
while the hole component depends on
\[
-\cos (k_F r) + \frac{\omega}{\sqrt{\Delta_0^2-\omega^2}} \sin (k_F r),
\]
basically showing a phase shift between electron and hole for strongly bound in-gap states ($\omega\approx 0$). Let us say a word of caution about using the above expression very close to the impurity where the Green's function, Eq. (\ref{GBCS}), fails as explained in section \ref{Numerical} and Ref.~\onlinecite{Mier2_2021}. 
For a more realistic non-local impurity potential, such as the one's of Ref.~\onlinecite{Flatte_1997b}, the particle and hole components can be different even at the impurity site and with zero non-magnetic scattering, $U_j=0$.

\subsection{Quantum phase transition}
A quantum phase transition (QPT) refers to a fundamental change in the ground state of a many-body system. To observe and study such transitions, we often track specific system parameters, such as the order parameter or gap, denoted as $\Delta$ in Eq. (\ref{Gap_realspace}) in our current context.

When a single magnetic impurity is present on a superconductor, the cooper pair binding energy is weakened and the emergence of in-gap quasiparticle states follows. As the magnetic interaction of the impurity with the superconducting electrons becomes prominent, the order parameter undergoes an abrupt change and a QPT is realized. The discontinuous change in the order parameter characterizes these transitions as first-order. In the present case, the QPT signals a new magnetic ground state~\cite{Salkola_1997,Sakurai_1970}.

\begin{figure}[t!]
    \begin{center}
    \includegraphics[width = 0.385\textwidth]{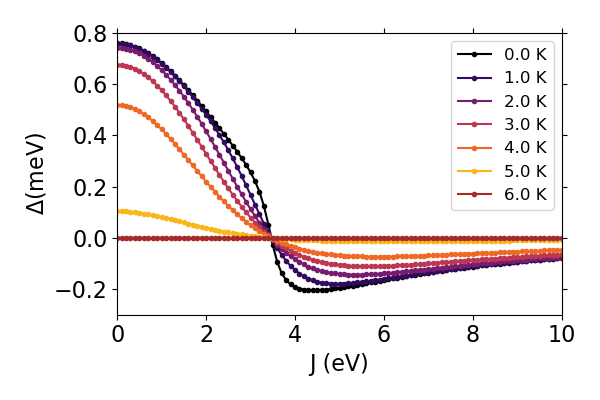}
    \end{center}   
    \caption{Order parameter, $\Delta$ as a function of exchange coupling strength, $J$, for a range of temperature values below $T_c$. The calculations shown here are for a single magnetic impurity on a 2D superconducting lattice. 
    These calculations have been performed without potential scattering, $U = 0$, or spin-orbit coupling, $\alpha = 0$. The used Fermi wave vector is $k_F = 0.183\, a_0^{-1} $. }
    \label{GapvsJ_allT}
\end{figure}

Figure~\ref{GapvsJ_allT} shows the order parameter as a function of effective impurity-sample interactions $J$ given in Eq. (\ref{Ham}). At zero temperature, a QPT takes place at a critical value of the exchange coupling, $J_c=3.46$ eV, when the gap at the impurity site becomes zero. As before, we have particularized to Bi$_2$Pd as superconductor. We have used the following parameters \cite{Herrera_2015}, lattice constant of 3.36 {\AA},  $k_F$ of 0.183 $a_0^{-1}$ and a Dynes parameter of 0.05 meV, which leads to a substantial smearing of the sharp transition near to the closing of the gap.
We used a Chromium-atom impurity, with spin S = 5/2. The impurity potential scattering, $U_i$, and  the surface's Rashba coupling have been neglected for Fig.~\ref{GapvsJ_allT}. As the temperature is increased, the gap of the pristine superconductor becomes smaller following Fig.~\ref{BCSgap}. However, the QPT can still be detected by the change of sign of the gap, even for temperatures close to the critical temperature.

\section{Several impurities}

The above  single impurity will produce an in-gap state with particle and hole components in the BdG formalism~\cite{YuLuh_1965,Shiba_1968,Rusinov_1969,Sakurai_1970}. The QPT that we just described for one impurity can be described in BdG terms as a crossing of the electron and hole characters due to the closing of the onsite gap~\cite{Sakurai_1970,Salkola_1997}. As the Kondo exchange interaction, $J$, of Hamiltonian, Eq. (\ref{Ham}), increases, the binding energy of the in-gap state increases. The BdG quasiparticles show that the particle and hole components approach in energy. When the gap closes, the energy of both components is the same (zero) and a crossing of the two components takes place leading to a crossing of the ground state with the first excited state of the system in a mean-field description~\cite{Sakurai_1970}.

\begin{figure}[t!]
    \begin{center}
    \includegraphics[width = 0.4\textwidth]{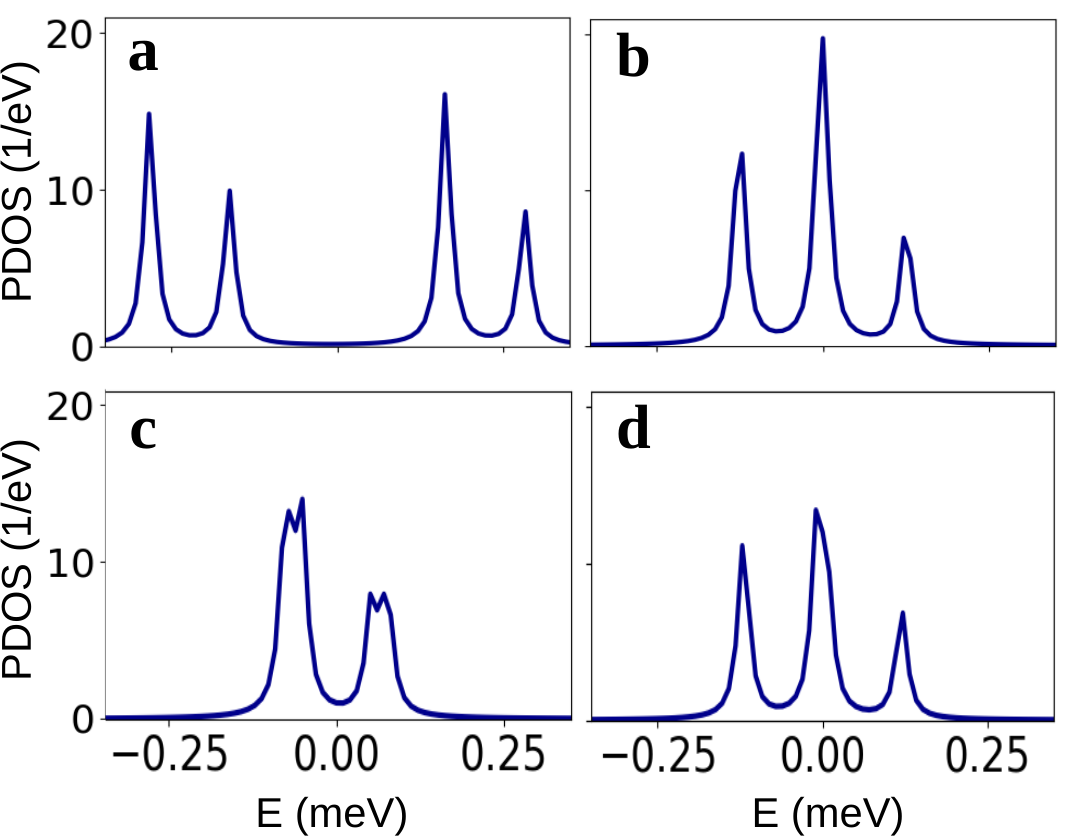}\end{center}  
    \caption{(a)-(d) shows the local density of states measured on one of the impurities of a magnetic dimer, displaying a sequence of in-gap states as the Kondo coupling strength, $J$, is changed to  (a) 2.5~eV, (b) 3.1~eV, (c) 3.4~eV and (d) 3.6~eV. The plots show zero-energy level crossings leading to the appearance of two quantum phase transitions for the case of a ferromagnetic dimer. The $J$-values at which the electron and hole parts of the shiba states cross each other are identified with critical $J$-points at which abrupt changes in the order parameter occurs (Fig.~\ref{spin_chain}). In this calculation for $k_F$ = 0.5 $a_0^{-1}$, the potential scattering is taken as $U$ = -5.5 eV, the spin is $S$ = 5/2,  and we do not include spin-orbit coupling, $\alpha$ = 0.}
    \label{LDOS_DIMER}
\end{figure}

\begin{figure}[t!]
    \begin{center}
    \hspace*{-0.7cm}
    \includegraphics[width = 0.5\textwidth]{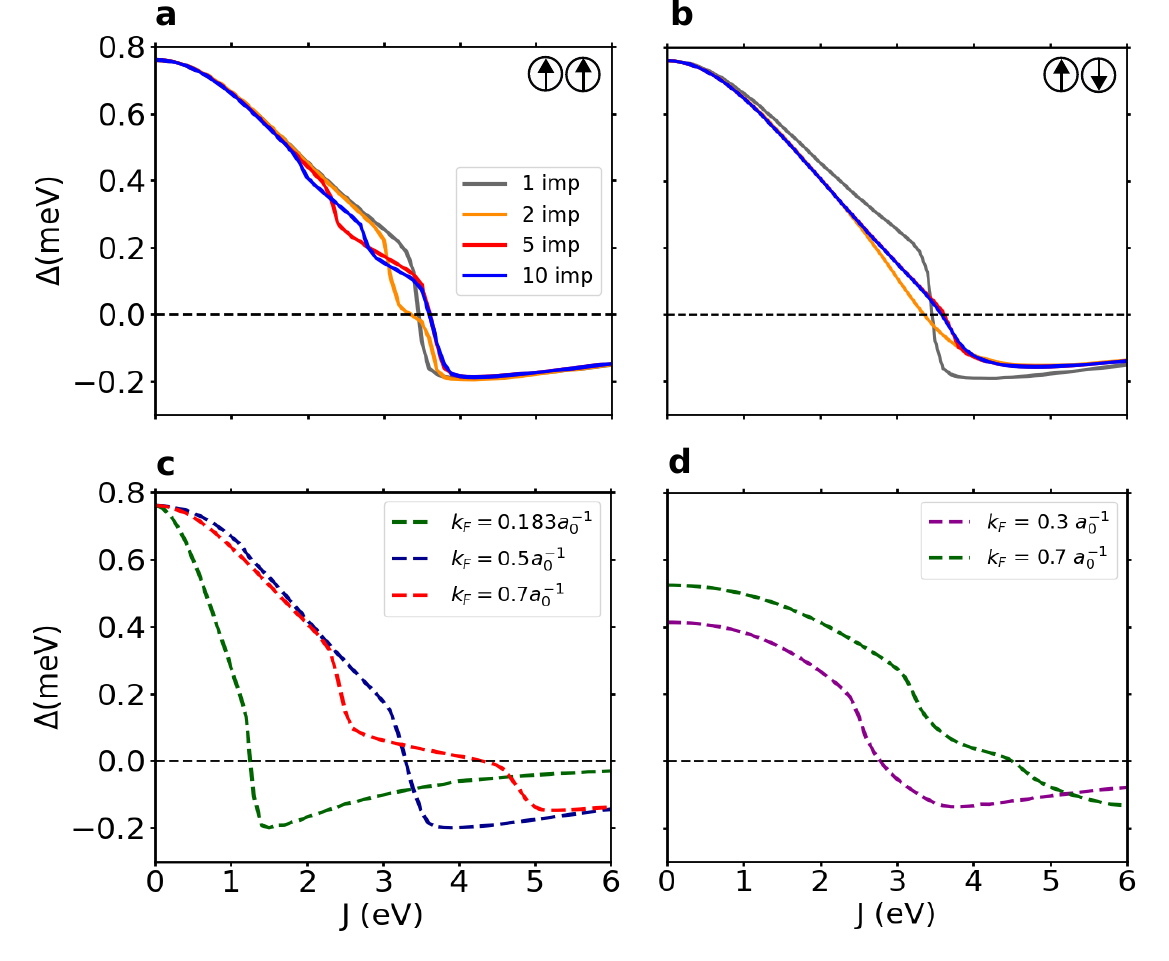}  
    \end{center}   
    \caption{Order parameter, $\Delta$ measured on the first atom of a finite spin-chain as a function of magnetic coupling strength, $J$, for impurities aligned in (a) ferromagnetic and (b) antiferromagnetic ordering. The spin-chains shown here comprise 1, 2, 5 and 10 atoms. Panels (c) and (d) show order parameter evaluated for an infinite-chain of ferromagnetic impurities. The abrupt changes in slope correspond to a first-order QPT. Depending on the number of atoms in the impurity chain, more than one QPT can be realized. In long enough chains, these abrupt changes smooth out into two QPT, one of which is identified as the topological one (where $\Delta$ crosses the zero). All calculations are performed without potential scattering ($U=0$) and without spin-orbit coupling ($\alpha$=0) except in (d) where $U$ is 5.5~eV and the Rashba coupling is $\alpha= 3$ eV-\AA. } 
    \label{spin_chain}
\end{figure}

When two impurities are present, a hybridization of the in-gap states can take place~\cite{Flatte_2000,Morr_2006,Meng_2015}. As a consequence, the in-gap state split in energy and the above picture becomes more complex. For ferromagnetically aligned states (or in general, a superconductor with a large spin-orbit coupling) the hybridization between in-gap states can be sizeable leading to crossings of the two electron and the two hole components for different $J$ values. This translates into two abrupt transition of the gap as a function of $J$. However, only one of the transitions leads to a closing of the gap. The gap, closes for the larger $J$, when both in-gap states have exchanged their electron-hole character.
This is clearly seen in Fig.~\ref{LDOS_DIMER}. There the local density of states (LDOS) at site $i$, $\rho_i (E)$, evaluated at a given quasi-particle energy $E$ is given by,
\begin{equation}
\rho_i (E) = -\frac{1}{\pi} Im [G_{i,i} (E)],
\end{equation}
where $G_{i,i} (E)$ is the local Green's function evaluated under one of the impurities. The zero of energy corresponds to the Fermi energy. There are four peaks corresponding to the two in-gap states. As before, we have used a complex energy, $\omega+i \Gamma$ using a Dynes parameter, $\Gamma$, of 0.01 meV this time. The imaginary value leads to an effective broadening of the density of states. As $J$ increases, the particle and hole peaks start reducing their energy difference until they cross. Each of the crossings  correspond to one of the two transitions in Fig.~\ref{spin_chain} for the dimer curve, in agreement with previous studies~\cite{Meng_2015}.

Figure~\ref{spin_chain} (a) shows the result of the self-consistent order parameter as a function of the Kondo exchange coupling, $J$, for 1, 2, 5 and 10 impurities located at an inter-impurity distance of $3.36$~\AA~ for an s-wave superconductor with $k_F=0.5~a_0^{-1}$ and $\Delta_0=0.76$~meV. We retrieve the previous results of a single QPT for a single impurity and two QPT for two impurities ferromagnetically coupled. As the number of impurity increases, more QPTs take place, leading to several values of $J_c$. The larger value of $J_c$ does not seem to increase for more than 5 impurities as we can corroborate by looking at the infinite-chain order parameter in Fig.~\ref{spin_chain} (c) that will be analyzed in the next section.

The antiferromagnetic spin chains, Fig.~\ref{spin_chain} (b), have a very different behavior. The transition does not seem to change its critical $J$ value, and it smooths out for an increasing number of atoms. This behavior can be understood considering the localized structure of the in-gap states for antiferromagnetic chains. The hybridization of the local in-gap states to form a band is more difficult leading to a small dispersion of the in-gap bands and to a more uniform character of the in-gap evolution.

\subsection{Infinite spin chains}

The infinite ferromagnetic spin chain can be calculated using Eq. (\ref{Gap_kspace}). The corresponding order parameters are plotted as a function of the exchange interaction $J$ in Fig.~\ref{spin_chain} (c). Contrary to the previous finite ferromagnetic chains, a very low number of transitions can be discerned. This has to do with the creation of bands of
in-gap states that have a very small number of gap closings \cite{Bjornson_2015}.
We have explored the order parameter as a function of $J$, for three different values of $k_F$ to probe low $k_F= 0.183 \; a_0^{-1}$, mid $k_F= 0.5 \; a_0^{-1}$ and high $k_F= 0.7 \; a_0^{-1}$ electronic densities. For low and mid densities, we find that there is a single QPT in the range of studied $J$ values. The lower the electronic density, the lower $J_c$ gets, we understand this by the increasing screening of the impurity's magnetic moment at larger electron densities, leading to larger values of $J_c$ to be able to produce an effect on the superconductor. 

At even larger electron densities ($k_F= 0.7 \; a_0^{-1}$), there are two clear transitions appearing, showing that the induced in-gap band structure matters and can give rise to complex behavior. 

Figure~\ref{spin_chain} (d) shows two cases for topological QPT (TPQT) in agreement
with previous calculations~\cite{Mier2_2021}, that indeed are very similar to the cases in (c), except for the presence of potential scattering and spin-orbit coupling.
The effect of spin-orbit coupling bridges the order-parameter behavior between the ferro- and antiferro-magnetically ordered chains. Thus, the transitions become smoother and closer
in values of $J$. At $J=0$ the two calculations show different $\Delta_0$ results because the self-consistent procedure has been tuned to reproduce the topological results of Ref.~\onlinecite{Mier2_2021} at the transition value. At low electronic densities, lower $k_F$, the spin-orbit coupling not only washes out the discontinuous jumps in the order parameter but also delays the occurrence of the QPT to a higher $J_c$ \cite{Bjornson_2016, Bjornson_2017}, however, for high $k_F$ regimes, the Rashba coupling is inefficient in changing the hybridization of the magnetic impurity with the electrons in the bulk superconductor and thus the results are unaltered.

\begin{figure}
    \centering
    \hspace*{-0.5cm}
    \includegraphics[width = 0.5\textwidth]{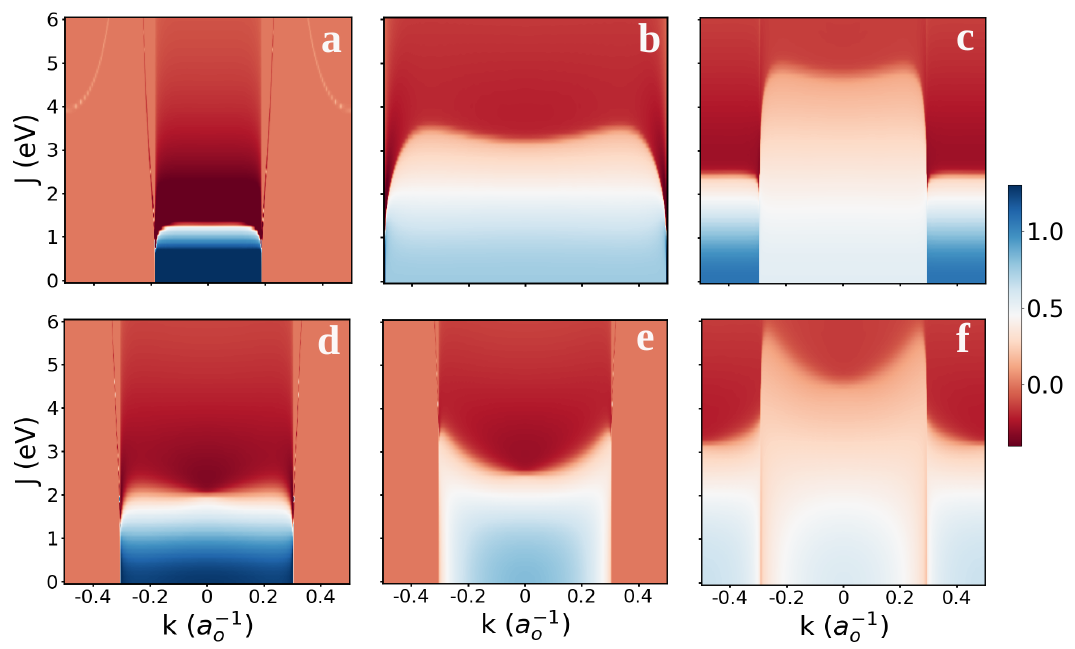}
    \caption{Plots showing the distribution of the order parameter, $\Delta_{k}$, in k-space as the Kondo exchange coupling strength, J is varied between 0 and 6~eV at different electronic densities represented by the Fermi wave vectors, $k_F$ taking values of (a) 0.183 $a_{0}^{-1}$, (b) 0.5 $a_{0}^{-1}$ (c) 0.7 $a_{0}^{-1}$, (d) 0.3 $a_0^{-1}$ while  the Rashba coupling and the scattering potential are absent. In the following cases, $k_F$ is (e) 0.3 $a_{0}^{-1}$, and (f) 0.7 $a_{0}^{-1}$ wherein the influence of Rashba coupling strength, $\alpha$ of 3 eV-\AA\ and potential scattering, U of 5.5 eV is considered. The color bar on the right denotes $\Delta_{k}$ measured in units of meV. Figures (a), (b) and (c) corresponds to the three curves shown in Fig.\ref{spin_chain}(c) while figures (e) and (f) correspond to the two curves of Fig.~\ref{spin_chain} (d).} 
    \label{2dmap}
\end{figure}

Figure~\ref{2dmap} shows the order parameter resolved in $k$-space as a function of the exchange interaction, $J$. 
For lower electron densities we find that there is not a large variation with $k$, reflecting the very local behavior of the gap. As a consequence, the gap closes for virtually all k-points at the same $J_c$, leading to a well defined single $J_c$. As $J$ increases, the gap does not seem to close
again. This is consistent with the single QPT found in Fig.~\ref{spin_chain} (c). 

As the electron density increases, the gap closes at different $J_c$ as $k$ changes, but the $J_c$'s remain relatively close to each other. As a consequence, there is basically a single effective $J_c$ with a small dispersion leading to a smoother transition as compared to the lower-density case. 

At higher densities, the folding of the bands due to a Fermi wave vector larger than the Brillouin-zone edge, $k_{BZ}$, leads to two distinct regions, $k > k_{BZ}$ and $k < k_{BZ}$ and thus, there are two well-defined $J_c$ values, seen in Figs.~\ref{spin_chain} (c) and (d) for $k_F=0.7\;a^{-1}_0$ as two abrupt changes in the slope of the $\Delta$ vs $J$ curves. 
More detailed information can be obtained from Fig.~\ref{Gap_k}. This figure permits us to analyze the behavior of $\Delta_k$ at high densities ($k_F=0.7\; a_0^{-1}$) as a function of $k$, for a few $J$ values around the critical values, $J_c$. $\Delta_k$ closes for all values of $J$ between $J = 3.2$ eV, and $J = 5.9$ eV, however the average order parameter $\Delta= 1/N_k \sum_k \Delta_k$ is zero for a unique $J=4.6$ eV. Figures \ref{spin_chain}, \ref{2dmap} and \ref{Gap_k} yield that there are two QPT for an infinite ferromagnetic spin chain at $k_F=0.7\; a_0^{-1}$ and the average gap only closes once.

\subsection{Topological quantum phase transition}

The cases in Fig.~\ref{spin_chain} (d) and Fig.~\ref{2dmap} (e) and (f) correspond to systems
that undergo TQPT as shown in Ref.~\onlinecite{Mier2_2021} where finite potential scattering and Rashba coupling are operative. As the exchange coupling $J$ is increased,
Ref.~\onlinecite{Mier2_2021} shows that there are two TQPT creating a topological zone between two trivial regions at larger and smaller $J$. Let us choose two different electronic densities in the phase space of Ref.~\onlinecite{Mier2_2021}, namely $k_F=0.3 \, a_{0}^{-1}$ and 0.7 $a_{0}^{-1}$. From Fig.~\ref{spin_chain} (d) we find two $J_c$ delimiting the topological region in $J$ for $k_F=0.7 \, a_{0}^{-1}$, however for $k_F=0.3 \, a_{0}^{-1}$ only one $J_c$ is clearly identified.

As before, Fig. \ref{2dmap} contains the information on the evolution of the order parameter. For the low-density case ($k_F=0.3 \, a_{0}^{-1}$) we find that there is a small interval of $J$ values between $-k_F$ and $+k_F$ where the gap becomes zero. The dispersion of these critical $J$ values with $k$ is small, leading to a single transition in Fig. \ref{spin_chain}, where the dispersion leads to a small broadening of the transition region, as we show in the preceding section.

\begin{figure}
    \centering
    \includegraphics[width = 0.3\textwidth]{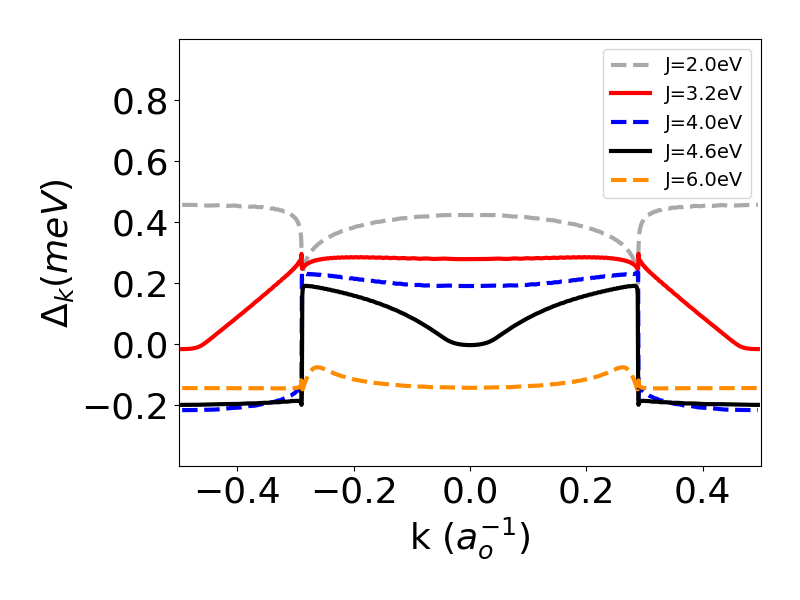}
    \caption{Order parameter as a function of wave-vectors, k for 5 different values of J such that the system is either in trival or topological region and at the junction of quantum phase transition. It is observed that first TQPT corresponds to closing the gap at two k-points. This is the curve corresponding to $J=3.2$ eV where $\Delta_k=0$  at the edges of the first Brillouin zone. For the second TQPT, the order parameter is zero at a single point $k = 0$, which corresponds to the curve for $J=4.6$ eV. Here, the two values of the critical exchange couplings, $J_c$, agree well with our non self-consistent results previously calculated in \cite{Mier2_2021}.The parameters used for this calculation corresponds to the data presented in Fig.~\ref{2dmap}(f).}
    \label{Gap_k}
\end{figure}

At higher densities ($k_F=0.7 \, a_{0}^{-1}$), we find two transitions due to the two different regions in $k$ space as discussed in the preceding section. To gain insight into the behavior of  the order parameter during these TQPT, we plotted the k-resolved order parameter, $\Delta_{k}$ as a function of k for different values of J corresponding to trivial and topological regions for large electron-density case, $k_F=0.7\;a_{0}^{-1}$. The results are shown in Fig.~\ref{Gap_k}. At $J=3.2$ eV, the value of $\Delta_{k}$ is zero only at the edges of the first Brillouin zone. As J is ramped, the system remains in this quantum phase until a second phase transition occurs at $J=4.6$ eV. This transition is identified when $\Delta_k = 0$ at $k=0$. For larger $J$ values, the system stays in the topologically trivial region and the gap does not close anymore as can be seen in Fig.~\ref{Gap_k} for $J=6$ eV. 

In brief, we can conclude that for large electronic densities, the first TQPT takes place at large $k$ ($J=3.2$ eV) and the second TPQT occurs at a smaller $k$ value ($J=4.6$ eV). 

\begin{figure*}
    \centering
    \includegraphics[width = 0.9\textwidth]{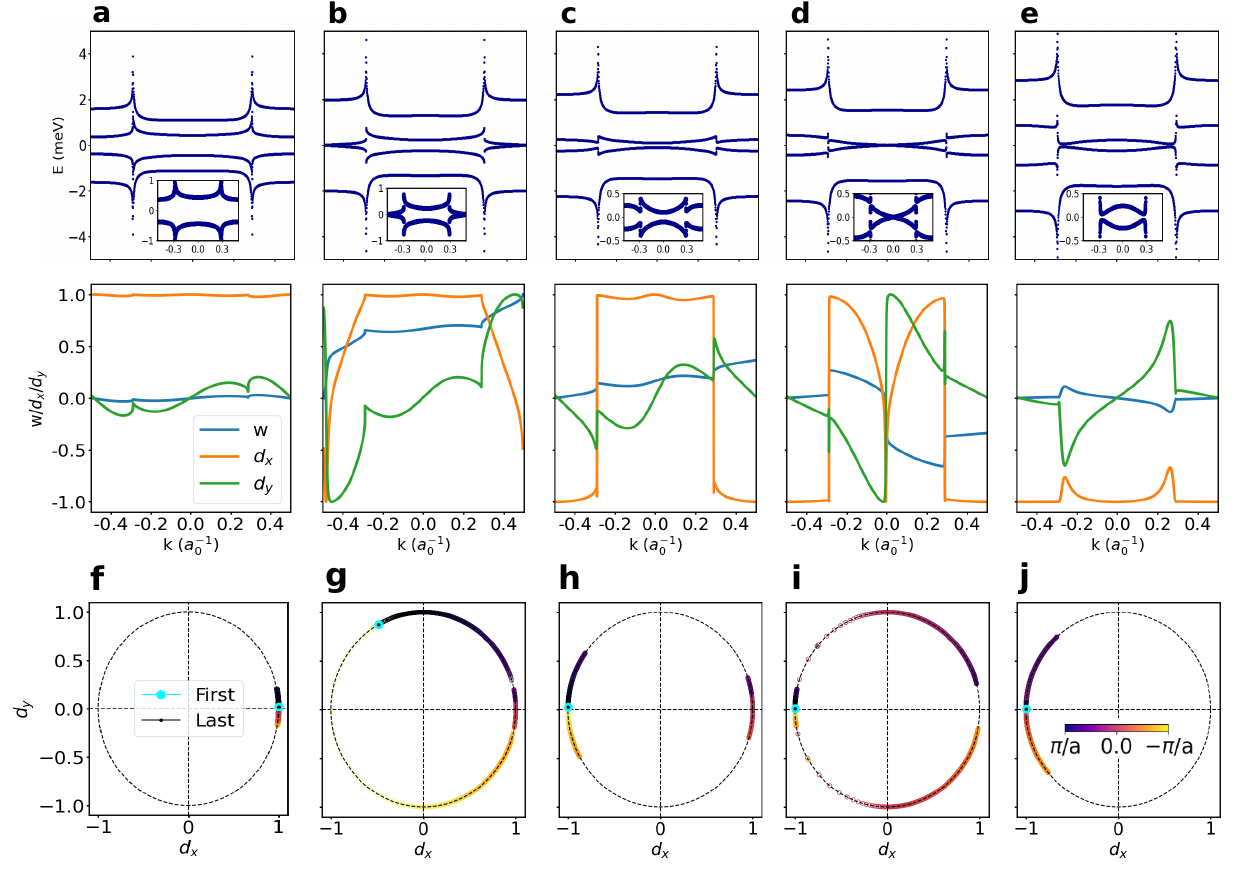}
    \caption{Energy band diagrams ((a)-(e) top) and evolution of winding vector $\vec{d}$ = ($d_x, d_y$) ((a)-(e) bottom) as a function of wave-vectors, k . The inset of the top set of figures simply shows a zoomed in cut-out of the bands near zero energy. From left to right, J = 2.0, 3.2, 4.0, 4.6 and 5.0 eV respectively. (f)-(j) illustrates the trajectory of normalized winding vector, $\vec{d}(k)$ as k is varied in the first Brillouin zone. The corresponding plots showcase the evolution of topological quantum phase transition as J is tuned through a range of values. The first TQPT is marked by a zero band gap  at the boundary of the Brillouin zone where the effect of band folding is observed. This is the onset of the topological region which is verified by a completion of one full turn of the winding vector, shown in (g). The second TQPT is observed when topological band closes at k=0. It can be observed that the winding vector already ceases to be 1 (g) and this leads to transition to the trivial region where the winding number goes back to 0 (j). A potential scattering of U = 5.5 eV, Rashba spin-orbit coupling of $\alpha$ =3 eV-{\AA} and fermi wave-vector, $k_F = 0.7 a_0^{-1}$ is used for this calculation.
    }
    \label{Bands}
\end{figure*}

The topological character of these transitions is revealed by studying the in-gap bands. Figure~\ref{Bands} shows the band structure and the winding number for the above example, Ref. \onlinecite{Mier2_2021}. The insets of Fig.~\ref{Bands}(a)-(d) clearly identifies closing and reopening of the topological band gap  at large $k$ for small $J$ and small $k$ for large $J$ as we just saw. This result agrees with the corresponding winding number calculation (bottom set of figures (a)-(d) and figures (f) - (j)).

The winding number is a topological invariant which is characterized by counting the number of complete turns in anticlockwise direction that the winding vector takes around zero. In Ref. \onlinecite{Mier2_2021} a complete account of the theory that we are using here is given. The winding vector, $\vec{d}$ is ($d_{x}, d_{y}$) that corresponds to the real and imaginary part of the Hamiltonian in the local basis set. The vector draws a closed trajectory in the first Brillouin zone. Depending on whether the zero is enclosed in the vector's trajectory, the topology will be trivial or not. The reason after this is that taking the zero out or in the vector's trajectory implies to make the Hamiltonian matrix elements zero and hence to close the band gap. Then, we can plot the evolution of ($d_{x}, d_{y}$) as a function of $k$ as done in the bottom set of Figs. \ref{Bands} (a)-(d) or we can plot $d_x$ vs $d_y$ as $k$ is changed, Figs. (f) - (j). Both sets of figures contain the same information, but it is easy to discern when the $d$-vector really turns around zero by comparing both sets. Furthermore, the lower-symmetry $Z_2$ invariant can be obtained from the values of $d_x$ at the Brillouin zone center and edges, see Ref.~\onlinecite{Mier2_2021}.

From these figures we obtain that indeed there are two $J_c$, the lower one closing the gap for $k$ values close to the Brillouin-zone edge, while the larger one closing the gap at the center of the Brillouin zone. And that the topological $Z_2$ invariant changes between the regions delimited by the $J_c$. The $d_x$ changes sign between Brillouin zone edge and center for $J$ values between the two $J_c$ showing the topological character
of the transition. The same information is confirmed by the winding number. This shows that between the two $J_c$ there is a topological non-trivial phase, while outside this region the phase is topologically trivial.

Equivalent results for the low-density case ($k_F=0.3 \, a_{0}^{-1}$) show that the gap only closes about $k=0$ in agreement with Fig. \ref{2dmap} (e), explaining that indeed, there is only one TQPT for this case at $J_c = 2.6$ eV. The actual evaluation of the order parameter and its self-consistent evaluation has permitted us to clearly show the TQPT. In Ref.~\onlinecite{Mier2_2021}, an attempt to plot the regions where the gap closed by following the in-gap band structure~\cite{Pientka_2013} was made, becoming increasingly difficult in certain areas of phase space. Here, we show that the self-consistent order parameter accurately gives this information in good agreement with the analysis of the topological invariant.

\section{Effects of the self-consistent procedure on the local density of states}

Closing the local gap should have important effects on the LDOS and hence on the result of STM studies of in-gap structure. Our calculations show that the LDOS before and after self-consistency are identical. Hence, the conclusions of the present study are that non-self-consistent calculations are enough to describe most experiments based on scanning tunneling microscope studies of the induced in-gap structure. Indeed, self-consistency has impact on the evaluation of the order parameter of superconducting gap, but it has no bearings on the actual crossings of in-gap states and hence on the in-gap bands. As a consequence the topology of BdG equations is the same with or without self-consistency.
    
All the above results have been obtained for wide-band superconductors, where the bandwidth of the metal phase is orders of magnitude larger than the superconducting gap as is the case for most s-wave superconductors. As a consequence, $\lambda_F$ is much shorter than the coherence length $\xi$. Since $\lambda_F$ is the scale of distances where the self-consistent gap changes, and $\xi$ is the usual superconducting length scale, the self-consistency does not change most of the superconductor, and does not affect its main properties. Examples are the above in-gap states, leading to DOS and in-gap band structures that are largerly unaffected during self-consistency.

Previous results in the literature \cite{Bjornson_2015,Bjornson_2017,Theiler_2019}
were computed in the case when $\lambda_F \sim \xi$ that corresponds to superconducting gaps in the order of magnitude of the normal-metal bandwidth.
We could not reach this very low electron densities, but the results from these works show that self-consistency qualitatively alter the in-gap structure and the topological character of the induced bands.

\section{Discussion and conclusions}

We have studied the self-consistent order parameter in the context of magnetic impurities in realistic bandwidth superconductors. The self consistency is computed by introducing the variation of the order parameter as a self-energy in the Dyson equation and calculating the complete Green's function of the system. This allows us to evaluate the new order parameter from the Green's function and its variation with respect to the previous iteration. This procedure is iterated until the variations in the order parameter are negligible. We have reproduced well-known BCS results such as the temperature behavior of the order parameter and rationalized the induced in-gap structure by a single impurity, including the spatial extension of the in-gap states and the temperature behavior of the QPT as  a function of the impurity-substrate coupling.

As we increase the number of impurities on the substrate, the order-parameter discontinuities give precise information on the appearance of QPT associated with the crossings of the chemical potential of the different in-gap states as the impurity-substrate magnetic interaction increases. For ferromagnetic spin chains of increasing number of impurities the discontinuities in the order parameter with the exchange interaction seem to yield a maximum number of QPTs. A careful study in $k$-space show that even if there is a large number of chemical potential crossings, the behavior of the QPTs is rather given by the order-parameter as a function of $k$-point. Thus, the order parameter contains very relevant information on the presence of QPTs.

Antiferromagnetic chains present a single QPT in short chains due to the localized character of the in-gap states and of the induced bands. When spin-orbit coupling is introduced, the situation becomes less clear and the transitions become very smooth, trending to a mixed behavior of the order parameter for ferromagnetic and antiferromagnetic spin chains.

In the presence of spin-orbit coupling, we can study the topological phases of the spin chain through the order parameter. In addition to understanding the phase space where QPTs are produced, studying the topological invariants at the same time as the self-consistent order parameter gives clear onsets of the different topological phases. Hence, the self-consistent order parameter becomes an interesting object to study the phase space for QPTs and particularly for TQPTs. 

For the present systems, we find fast convergence and virtually no impact of self-consistency on the evaluation of the density of states and on the topological character of the induced in-gap bands. This is a consequence of the short-range changes in the order parameter that die within Fermi-length scale, $\lambda_F$ from the impurity. However, the superconducting properties are set within the coherence length, $\xi$, which is typically orders of magnitude larger in realistic s-wave superconductors. Thus, non-self-consistent calculations in large-bandwidth superconductors give accurate results on the induced in-gap structure contrary to the case of small-bandwith superconductors~\cite{Bjornson_2015,Bjornson_2017,Theiler_2019}. 

In summary, computing the order-parameter is a small computational overhead that  gives valuable information on QPT, temperature dependence of the superconducting properties and the topological phase space of the in-gap structure.

\begin{acknowledgments}
We are pleased to thank Dr. Cristina Mier for discussions and assistance in the early stages of this work. Special thanks to Prof. Michael Flatt\'e for insightful discussions on the topic. This work was supported by the project  PID2021-127917NB-I00 funded by MCIN/AEI/10.13039/501100011033, QUAN-000021-01 funded by the Gipuzkoa Provincial Council and IT-1527-22 funded by the Basque Government. 
\end{acknowledgments}

\appendix
\section{Calculation of the e-h coupling constant}
\label{AppendixV}

\begin{figure}[h!]
    \begin{center}
    \includegraphics[width = 0.4\textwidth]{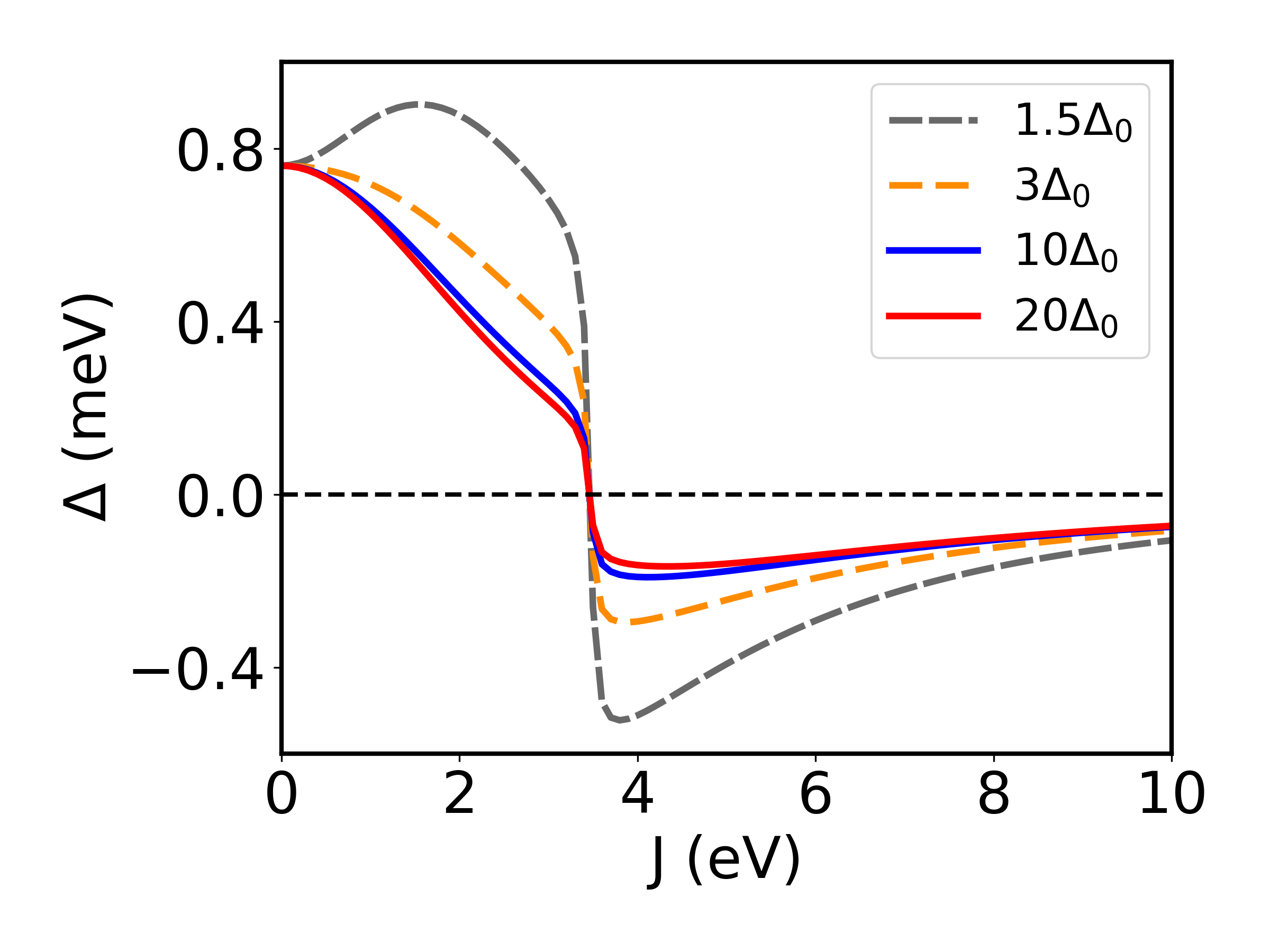}
    \end{center}  
    \caption{The self-consistent gap as a function of the impurity's Kondo exchange interaction for four different cutoffs $\hbar \omega_c=1.5\Delta_0, 3 \Delta_0$, $10 \Delta_0$ and $20\Delta_{0}$, where $\Delta_0=\Delta (J=0)=0.76$ meV. A low cutoff leads to a qualitatively wrong behavior of the order parameter, although the critical Kondo exchange interaction, $J_c$, does not change. 
    }
    \label{cutoff}
\end{figure}

The calculation of Eq. (\ref{GapSelf}) is difficult because despite the natural cutoff of the Debye frequency, $\omega_D$, the superconducting gap, $\Delta$ is still  one or two orders of magnitude smaller than  $\omega_D$, for typical realistic-bandwidth BCS superconductors. Here, we assume a second cutoff, $\omega_c$ that is smaller than $\omega_D$, but large enough to avoid problems in the self-consistency. If the cutoff is too small, the self-consistent gap shows an increase in value as $J$ increases, which disappears if the cutoff is increased, Fig.~\ref{cutoff}. 

The procedure to attain self-consistency~\cite{Flatte_1997a,Flatte_1997b} is to include $\delta \Delta$ as a self-energy in Dyson's equation, Eq. (\ref{Dyson2}), where $\delta \Delta$ is the change in the gap between self-consistency iterations. The integral of Eq. (\ref{GapSelf}) between both cutoffs, $\omega_c$ and $\omega_D$ is assumed to be a constant, because for large-enough energy, we expect to recover the normal-metal electronic structure.   The advantage is that we do not need to evaluate the integrals between cutoffs in  $\delta \Delta$.

In order to evaluate $\tilde{V}$ of Eq. (\ref{GapSelf}), we use that $\omega_c \gg {\Delta}_0$ such that
\begin{equation}
Im(G^{1,4}_{i,i}(\omega)-G^{2,3}_{i.i}(\omega)) = -Re(\frac{2\pi N_0 \Delta_0}{\sqrt{(\omega+i\gamma)^2-\Delta_0^2}}) 
\end{equation}
for temperatures small enough to be in the superconducting phase ($k_B T \ll \hbar \omega_c$) and for small Dynes parameters ($\gamma \ll \Delta$), we have that the correction to the gap because the cutoff is actually $\omega_D$ is
\begin{eqnarray}
\delta\Delta_i &=& \tilde{V}
\int^{-\hbar \omega_c}_{-\hbar \omega_D}
f(\omega) Im \{G^{1,4}_{i,i}(\omega)-G^{2,3}_{i.i}(\omega)\} d\omega \nonumber \\
&\approx& -\tilde{V} \; 2 \pi N_0 {\Delta}_0 \; ln \left ( \frac{\omega_c}{\omega_D} \right ).
\label{GapCorr}
\end{eqnarray}
Given the qualitative nature of the present study, we have ignored the correction and assumed that is partially
taken care of by the actual value of $\tilde{V}$, because we obtain $\tilde{V}$ from Eq. (\ref{GapSelf}),
for the clean-superconductor gap. However, the actual
critical values of the QPT will depend on the values taken for $\omega_c$ and $\omega_D$. All calculations
have been performed for $\omega_c=10 \, {\Delta}_0$, where the behavior of ${\Delta}_0$ is consistent with the results of previous studies~\cite{Salkola_1997,Flatte_1997b,Karan_2022} as can be seen in  Fig.~\ref{cutoff}.

\section{Role of non-magnetic potential scattering}

\begin{figure}[h!]
    \centering
    \includegraphics[width = 0.45\textwidth]{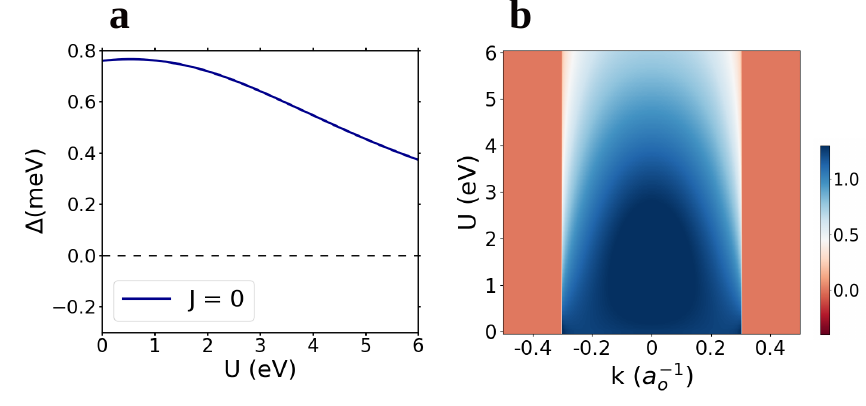}
    \caption{(a) Plot showing impact of non-magnetic coupling strength, U on the superconducting order parameter, $\Delta$. (b) 2d map shows the distribution of $\Delta_{k}$ in k-space as  U is ramped from 0 to 6 eV. Color bar corresponds to $\Delta$ given in meV units. The Rashba spin-orbit coupling is fixed to 3~eV-{\AA}, $k_F$ = 0.3 $a_0^{-1}$ and Kondo exchange coupling, $J$, is absent in this calculation. }
    \label{fig:GapvsU}
\end{figure}

According to Anderson's theorem, non-magnetic impurities do not alter s-wave superconductivity \cite{Anderson_1959}. However,  a non-magnetic impurity does have an effect although it cannot produce in-gap states \cite{Soda}.
Within our theory, it is clear that  the magnetic part is not the only one to contribute to the modification of the order parameter. Indeed, the non-magnetic coupling can have a significant contribution. Figure \ref{fig:GapvsU} shows that there is a reduction of about 48\% in the superconducting order parameter as the coupling strength is tuned from 0 up to 6 eV. Our results are in good agreement with the observations previously reported by Flatte \& Byers in Ref. \onlinecite{Flatte_1997}.

\bibliography{references}

\end{document}